\documentclass[12pt]{article}
\usepackage{epsf}
\newcommand{\gtrsim}{ \mathop{}_{\textstyle \sim}^{\textstyle >} }
\newcommand{\lesssim}{ \mathop{}_{\textstyle \sim}^{\textstyle <} }

\title{\begin{flushright}
\small{RESCEU-5/01} \\
\small{hep-ph/0105134} 
\end{flushright}
\vspace{2cm}
\Large
\bf Adiabatic and isocurvature fluctuations of Affleck-Dine field in
D-term inflation model}

\author{Masahiro Kawasaki  and Fuminobu Takahashi \\
\it{Research Center for the Early Universe (RESCEU),}\\ 
\it{School of Science,
University of Tokyo} }

\def\ds{\displaystyle}

\begin{document}

\maketitle
\vspace{3cm}
\begin{abstract}
We reconsider fluctuations of Affleck-Dine (AD) field in a D-term
inflation model. Contrary to the previous analysis,
we find that the spectrum of the adiabatic fluctuations is almost
scale invariant even if the AD field has a large initial value.
Furthermore, we study the isocurvature fluctuations of the AD field
and estimate the ratio of the isocurvature to adiabatic power
spectrum. The dynamics of the inflaton and AD fields sets the
upper bound for the value of the AD field, leading to a lower limit
for isocurvature perturbation. It is shown that the recent Cosmic
Microwave Background data give a constraint on the D-term inflation
and the AD field.
\end{abstract}
\clearpage
\section{Introduction}

Minimal Supersymmetric Standard Model (MSSM) have many flat
directions, along which there are no classical potentials. 
The flat directions are only
lifted by soft SUSY breaking mass terms and non-renormalizable
terms. Such flat directions have  several interesting consequences in
cosmology such as baryogenesis and Q-ball formation. In Affleck-Dine
mechanism for baryogenesis, a complex field along a flat direction (AD
field) has a large field value during inflation. After inflation the AD
field  starts to oscillate when its mass becomes larger than the Hubble
parameter $H$. At that time the AD field obtains velocity in the phase
direction because of the baryon number violating operator, 
and the baryon number is generated very efficiently.

For inflation models in which the accelerated cosmic expansion is caused
by a F-term potential, the AD field generally obtains an effective mass of
the order of $H$ through supergravity effects during inflation. In this
case, the expectation value of the AD field $\Phi$ is determined by the
balance  between mass term $c H^2 |\Phi|^2$ and the non-renormalizable
 term  where $c$ is order one constant. If the mass
term is positive ($c > 0$) as expected from minimal supergravity,
the potential takes minimum at $\Phi = 0$ and the AD mechanism does not
work. Thus the supergravity effects must lead to negative mass term (i.
e. $c < 0$) for baryogenesis in F-term inflation models.

Supersymmetric inflation models are also constructed with use of D-term
potential. In the D-term inflation models supergravity effects do not
induce $O(H^2)$ mass terms and the AD field has only soft SUSY breaking
mass term of the order of weak scale. Thus, the AD field can have a 
large expectation
value during inflation, which makes the AD baryogenesis work.

Since the effective masses of the AD field in D-term inflation models are
much smaller than the Hubble parameter during inflation, the AD field
has  large fluctuations which may give some contribution to the density
fluctuations of the universe. In the previous work~\cite{McE,McE2},  the
fluctuations of AD field was considered and it was pointed out that the
the fluctuations of the AD field change the spectral index of the
adiabatic fluctuations produced in the D-term inflation, from which the
expectation value of the AD field is constrained. 
The authors in Refs.~\cite{
McE,McE2} also pointed out that the small expectation value of the AD
field  during inflation leads to lager isocurvature (baryon)
perturbations which are induced by the fluctuations in the phase
direction of the AD field. However, their estimation of the adiabatic
fluctuations was too naive, and one needs careful treatment for the 
multi-field dynamics.

In this paper, we reconsider the AD field fluctuations in the D-term
inflation model. We find that the spectrum of the adiabatic fluctuations
is almost scale invariant even if the AD field has a large initial
value. Furthermore, we study the isocurvature fluctuations of the AD
field and estimate the ratio of the isocurvature to adiabatic power
spectrum.  It is shown that the recent Cosmic Microwave Background (CMB)
data give a constraint on the model parameters of D-term inflation and
the AD field.

\section{Dynamics of AD and inflaton fields}

Let us discuss the dynamics of the AD and inflaton fields in the D-term
inflation model.  The flat direction corresponding to the AD field can
be lifted by soft SUSY breaking mass term and the non-renormalizable
term coming from the superpotential given by
\begin{equation}
    W(\Phi) =  \frac{\lambda \Phi^d}{d M^{d-3}},
\end{equation}
where $M \equiv \frac{1}{\sqrt{8 \pi G}}$ is the reduced Planck mass,
$\lambda$ is an $O(1)$ coupling constant, and $\Phi \equiv \frac{\phi \,e^{
i \theta}}{\sqrt{2}}$ is the AD field.
Then, the potential is written as ~\cite{Dine}
\begin{equation}
    \label{eq:adp}
    V(\phi) =  m^2 |\Phi|^2 + \left(\frac{a m_{3/2} \Phi^{d}}{d M^{d-3}}+
        h.c. \right)+
    \frac{\lambda^{2} \left| \Phi \right|^{2(d-1)}}{M^{2(d-3)}},
\end{equation}
where $m$ is the soft mass, $m_{3/2}$ is the mass of gravitino, and
$a$ is an $O(1)$ complex constant.
The dimension $d$ is an even
number, and the cases of $d=4, 6 $ are considered in this paper. Note
that the AD scalar $\phi$ should be less than $O(M)$, otherwise the
potential cannot be described as Eq. (\ref{eq:adp})

The tree-level scalar potential for D-term inflation is given by \cite{
Dterm}
\begin{equation}
     V = \left | \kappa \right | ^{2}
     \left ( \left | \psi_{+} \psi_{-} \right |^{2}+
     \left |  S\psi_{+}  \right | ^{2}+\left |  S\psi_{-}  \right | ^{2}
     \right) +
     \frac{g^{2}}{2} \left (\left | \psi_{+} \right |^{2}
     -\left | \psi_{-} \right |^{2} + \xi^{2} \right)^{2},
\end{equation}
where $\kappa$ is the coupling constant of the interaction between
$S$ and $\psi_{\pm}$,  $\xi^{2}$ is the Fayet-Illiopoulos term, and
$g$ is the $U(1)_{FI}$ gauge coupling. Though the global minimum
of this potential is at $S=\psi_{+}=0$, $|\psi_{-}|=\xi$,  the local
minimum is at $\psi_{+}=\psi_{-}=0$ for $|S |> S_{c} \equiv
g \xi/\kappa$.  We can take $S$ real using the $U(1)_{FI}$ phase
rotation.  With use of  $\sigma \equiv  \sqrt{2}S$, for $\sigma >
\sigma_{c} = \sqrt{2} g \xi/\kappa$, the potential of $\sigma$ up
to a 1-loop correction  is given by~\cite{Dterm}
\begin{equation}
      V(\sigma)=V_{0}+\frac{g^{4} \xi^{4}}{16 \pi^{2}}
      \ln{\left(\frac{\sigma}{Q} \right)},
\end{equation}
where $V_{0}=\frac{g^{2} \xi^{4}}{2}$, and $Q = \sigma_{c}$ is a
renormalization point. If the potential is dominated by
$V(\sigma)$, $\sigma$ is related to the number of e-folds (
of inflation) $N$,
\begin{equation}
      \label{eq:efolding}
      \sigma \simeq \frac{g M}{2 \pi} \sqrt{N}.
\end{equation}
Assuming that the present scale crosses the Hubble radius at $N=50$
during inflation, the COBE normalization ($V^{\frac{3}{2}}/V'=5.3
\times
10^{-4}$) fixes
\begin{equation}
      \xi = 7.05\times 10^{15}~{\rm GeV}.
\end{equation}
%%
%With the gently-sloping potential, $\sigma$ always slowly rolls until
%$\sigma$ reaches $\sigma_{c}$, even when $V(\phi)$ is dominant
% (see below).

If $\phi$ starts with $\phi_{i} \sim O(M)$, the potential is dominated
by the F-term of the AD field. Then there arises $O(H^2)$ mass term for
the inflaton due to supergravity effects, and it rapidly rolls down to its true minimum. 
Therefore inflation does not occur at all in this case.
In order to have a successful inflationary model, $\phi_{i}$ must be less
than $\phi_{c}$, which is given by
\begin{equation}
\phi_{c} = \sqrt{2} \left(\frac{g}{\sqrt{2} \lambda} 
\xi^{2} M^{d-3} \right)^{\frac{1}{d-1}}.
\end{equation}
For $\phi_{i} \leq \phi_{c}$, the universe is dominated by the D-term of the
inflaton field.\footnote{
Precisely speaking, the inflation can occur even if  $\phi_{i} \gtrsim
\phi_{c}$ (say, $\phi_{i}=1.5 \phi_{c}$, $\sigma_{i}=M$).
In any case, the AD field value is set below $\phi_{sr}$  along
the same argument, once the D-term driving inflation begins.
}
Thus the D-term driving inflation accounts for
the necessary $O(50)$ e-folds. This gives $\sigma_{i} =O(g M)$.
Then it is easy to show that the slow roll condition for $\sigma$ is satisfied,
\begin{equation}
     \epsilon_{\sigma} \ll |\eta_{\sigma}| \simeq M^{2}
     \frac{V(\sigma)''}{V(\sigma)}
     \simeq \frac{1}{2N} \ll 1,
\end{equation}
where $\epsilon_{\sigma}$ and $\eta_{\sigma}$ are slow roll
parameters~\cite{Lyth}. Once the inflation begins, the AD field
rapidly oscillates with a decreasing amplitude.
When the amplitude of the oscillation
becomes  small enough ($\phi \lesssim \phi_{sr}$), $\phi$ begins to
slow-roll. $\phi_{sr}$ is obtained by solving $\eta_{\phi}(\phi_{sr})
= M^2\frac{V(\phi)''}{V(\sigma)}
\simeq 1$:
\begin{eqnarray}
      \phi_{sr} &=& \left(
      \frac{2^{d-1}M^{2d-8}V_{0} }{(2d-2)(2d-3) \lambda^{2}}
      \right)^{\frac{1}{2d-4}} , \\
      &=&\left\{
      \begin{array}{ll}
             \ds{ \left(\frac{2}{15} \right)^{\frac{1}{4}}
             \sqrt{\frac{g}{\lambda}} \xi} & \ds{d=4} \\
             \ds{\left(\frac{8}{45} \right)^{\frac{1}{8}}
             \left(\frac{g}{\lambda} \right) ^{\frac{1}{4}}
             \sqrt{\xi M}} & \ds{d=6}
      \end{array}
      \right..
\end{eqnarray}
Therefore, the expectation value of the AD field $\phi$ at the horizon exit
of the present scale is generally less than $O(\phi_{sr})$. Thus, we have
an upper limit to $\phi$,
\begin{equation}
       \label{eq:upper}
       \phi \lesssim \phi_{sr}.
\end{equation}
Note that this upper limit to the AD field value is totally due to its
dynamical property and the requirement that an inflation should occur,
not any observational constraint.

\section{Adiabatic Fluctuations}

\begin{figure}[htbp]
\centering
\hspace*{-7mm}
\leavevmode\epsfysize=14cm \epsfbox{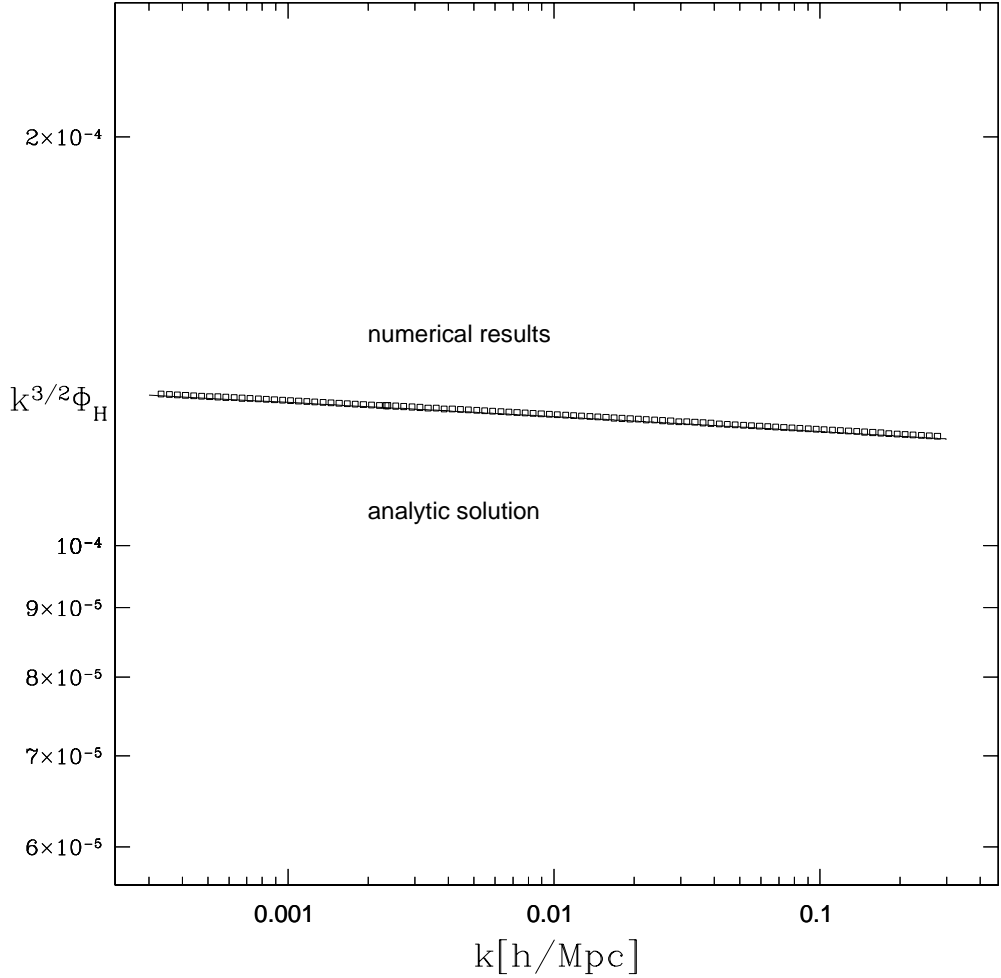}\\[2mm]
%\begin{center}
%\leavevmode\psfig{figure=spectrum.ps,width=16cm}
%\end{center}
\vspace{-1mm}
\caption[fig-ho]{\label{fig:spectrum}The analytic and numerical
results for the primordial spectra. We have taken $d=6$,
  $g=\lambda=\kappa=0.1$,
$\sigma_{i}=0.114M$, $\phi_{i}=0.0278M$ in numerical calculation.}
\end{figure}

Next, we calculate the fluctuations of the AD field $\phi$ and
inflaton $\sigma$.  According to Ref.~\cite{Sta}, the gauge-invariant 
curvature perturbation $\Phi_{H}$ 
is given by
\begin{equation}
      \label{eq:BPH}
      \Phi_{H}  = -C_{1} \frac{\dot{H}}{H} + \frac{C_{3}}{3 V_{total}^{2}}
     \left( V'(\sigma)^{2} V(\phi)-V(\sigma) V'(\phi)^{2} \right),
\end{equation}
where\footnote{%%
     As mentioned in \cite{doubleD},
     it is not so obvious how to distribute the constant $V_{0}$
     into $V(\sigma)$ and $V(\phi)$ in Eq. (\ref{eq:const}). We have
     checked Eq. (\ref{eq:const}) by numerical calculation as shown
     in Fig.\ref{fig:spectrum}. 
     }
\begin{equation}
      \label{eq:const}
      C_{1}=\frac{H}{V_{total}}
      \left( V(\sigma) \frac{\delta \sigma}{\dot{\sigma}}
     +V(\phi) \frac{\delta \phi}{\dot{\phi}} \right),
\end{equation}
is the growing adiabatic mode, and
\begin{equation}
       C_{3} = \frac{1}{2 H} \left(\frac{\delta \sigma}{\dot{\sigma}}
       -\frac{\delta \phi}{\dot{\phi}} \right),
\end{equation}
is the isocurvature mode. For $\phi < \phi_{sr}$, $V_{total}$ is
dominated by $V_{0}$, and $\dot{\sigma} \leq \dot{\phi}$.
Hence $C_{1} \simeq H\frac{\delta \sigma}
{\dot{\sigma}}$. In other words, the main contribution to the
adiabatic fluctuation comes from the inflaton. Therefore the primordial
spectrum is almost scale-invariant as usual. With this approximation,
the primordial spectrum is given by
\begin{equation}
       \label{eq:analytic}
       k^{\frac{3}{2}} \Phi_{H} (k) \simeq \frac{\sqrt{12}\pi}{5}
      \left(\frac{\xi}{M} \right)^{2}
      \sqrt{50-\ln{\frac{k}{k_{0}}}} ,
\end{equation}
where $k_{0} ^{-1} \equiv 3000 h^{-1}$Mpc.

In order to check the above estimate,  
we also solve the evolution of zero modes and
fluctuations of the AD and inflaton fields and the gauge-invariant 
curvature perturbation $\Phi_H$
by  numerical calculation, following Ref.\cite{BB}.
We present the evolution equations in
synchronous gauge as \cite{BB}
\begin{eqnarray}
\kappa & \equiv & \left(\frac{k^{2}}{a^{2}}\mathcal{R}-4 \pi G \delta \rho
\right) /H, \\
\delta \rho_{com}&\equiv& \delta \rho-3H \Psi,
\end{eqnarray}
\begin{eqnarray}
\dot{\mathcal{R}} &=& 4 \pi G \Psi, \\
\delta \dot{\rho}_{com} &=& -3 H\left(1+\frac{1}{2} (1+\omega) \right)
\delta \rho_{com}+\frac{k^{2}}{a^{2}} \left(\frac{(\rho+p)\mathcal{R}}{H}+
\Psi \right), \\
\dot{\Psi}&=&-3H \Psi-\delta p, \\
\delta\dot{p} &=& \frac{1}{3} \delta \rho_{com}+H \Psi +\frac{2}{3}
\sum_{j} \left(\dot{\phi_{j}} \delta \dot{\phi_{j}}-
2\frac{\partial V_{total}}{\partial \phi_{j}} \delta \phi_{j}\right),\\
\delta \ddot{\phi_{j}}&+&3H \delta \dot{\phi_{j}}+\frac{k^{2}}{a^{2}}
\delta \phi_{j}+ \sum_{i} \frac{\partial^{2}V_{total}}{\partial \phi_{j}
\partial \phi_{i}} \delta \phi_{i}-\kappa \dot{\phi_{j}}=0,
\end{eqnarray}
where $a$ is the scale factor,
$\phi_{1(2)}=\sigma \,(\phi)$, $\mathcal{R}$ is the curvature
perturbation, $\Psi \equiv
-\Sigma_{j}\dot{\phi_{j}} \delta \phi_{j}$ is the total momentum
current potential, $p$ and $\rho$ are the total pressure and energy
density, and $\omega \equiv \frac{p}{\rho}$.
The gauge-invariant 
curvature perturbation $\Phi_{H}$ is given by
\begin{equation}
\Phi_{H} = 4 \pi G \frac{a^{2}}{k^{2}} \delta \rho_{com}.
\end{equation}
We have found that the above  analytic
solution Eq.(\ref{eq:analytic}) agrees well with the numerical results.
These two primordial
  spectra are plotted in Fig.\ref{fig:spectrum}.
  From Fig.\ref{fig:spectrum}, one can see that the analytic solution (the
solid line) agrees with the numerical results very well, which support
the validity of the approximation used above.

\begin{figure}[htbp]
     \centering
     \hspace*{-7mm}
     \leavevmode\epsfysize=14cm \epsfbox{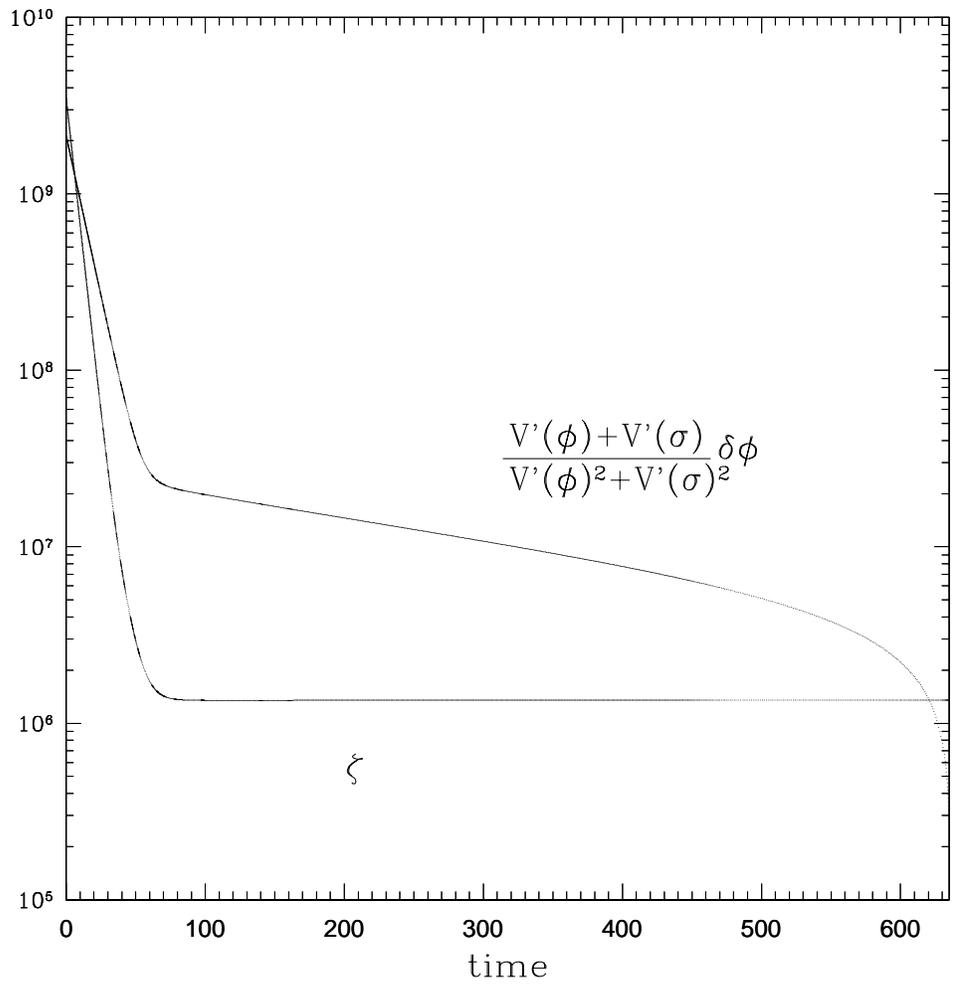}\\[2mm]
     \vspace{-1mm}
    \caption[fig-ho]{\label{fig:zeta}
         The time evolutions of the gauge invariant $\zeta$ and
         $\frac{V'(\phi)+V'(\sigma)}{V'(\phi)^{2}+V'(\sigma)^{2}} \delta \phi$
         during inflation.}
\end{figure}

The spectral index is given by
\begin{equation}
      \label{eq:index}
       n=1+2\frac{d \ln{k^{\frac{3}{2}} \Phi_{H}(k)}}{d\ln{k}}.
%    \frac{d\ln {T(k)}}{d \ln {k}},
\end{equation}
Substituting Eq.(\ref{eq:BPH}) ( or Eq.(\ref{eq:analytic})) into Eq.
(\ref{eq:index}),  the spectral index $n_{\rm COBE}$ on COBE scales
\begin{eqnarray}
       n_{\rm COBE} &\simeq &1-3M^{2}
       \frac{V'(\sigma)^{2}}{V_{0}^{2}}
      +2M^{2}\frac{V''(\sigma)}{V_{0}}-M^{2}
       \frac{V'(\phi)^{2}}{V_{0}^{2}}, \\
       &\simeq & 1+2M^{2}\frac{V''(\sigma)}{V_{0}}, \\
       & =& 1-\frac{1}{N_{COBE}} \simeq 0.98.
\end{eqnarray}
We have also obtained $n_{\rm COBE} = 0.98$ from our numerical
calculation. Note that $n_{\rm COBE}=1.2 \pm 0.3$ is  implied by
present CMB observations  and hence the CMB observation does not
restrict any parameters in this model as opposed to the result
in Ref.~\cite{McE2}. This discrepancy  is due to the incorrect estimation
of  the gauge invariant $\zeta$ in Ref.~\cite{McE2}  where $\zeta$ is
given by
\begin{equation}
      \label{eq:wrong}
      \zeta = \frac{\delta \rho}{\rho+p} \propto
      \frac{V'(\phi)+V'(\sigma)}{V'(\phi)^{2}+V'(\sigma)^{2}}
      \delta \phi.
\end{equation}
On the other hand, the accurate form of $\zeta$ is given by
\begin{equation}
      \zeta=\frac{\Delta_{g}}{1+ \omega}
      -\frac{16 \pi G a^{2} p}{(1+\omega)k^{2}} \Pi,
\end{equation}
where $\Delta_{g}\equiv \delta+3(1+\omega) \mathcal{R}$, $\delta \equiv
\delta \rho/\rho$, and $\Pi$ is an anisotropic stress
perturbation.  Hence only if we take the flat slicing and an anisotropic
stress perturbation can be neglected, then the first equality in
Eq.~(\ref{eq:wrong})  is satisfied. It is also worth noting that the
dynamics of the perturbed system cannot be described by one
equation for $\zeta$ when more than one scalar field are
involved~\cite{pps,nakano}.  In addition,  the expression (\ref {eq:wrong}) is
based on $\delta \sigma= \delta \phi$. However, $\delta \sigma$ and
$\delta \phi$ are classical random quantities,
and  the correct  expression is $ \langle(\delta \sigma)^2\rangle =
\langle (\delta \phi)^2\rangle$, where $\langle\cdots\rangle$ means
ensemble average.   The results of numerical calculation
for the time evolutions of $\zeta$ and
$\frac{V'(\phi)+V'(\sigma)}{V'(\phi)^{2}+V'(\sigma)^{2}} \delta \phi$
during inflation are plotted in Fig.\ref{fig:zeta}. As seen in
Fig.\ref{fig:zeta}, the two quantities are not proportional to each
other.
 
\section{Isocurvature Fluctuations}

During inflation, the fluctuation of the phase $\theta$ of  the AD field is
given by
\begin{equation}
      \delta \theta =\frac{H}{\sqrt{2 k^{3}} \phi}.
\end{equation}
After baryogenesis by the  AD mechanism, the fluctuation of $\theta$
becomes baryonic isocurvature perturbation~\cite{astro}. According to
Ref.~\cite{McE2}, the baryon number density $n_{B}$ is proportional
to $\sin {2\theta}$. Thus the isocurvature fluctuation with comoving
wavenumber $k$ is given by
\begin{eqnarray}
       \delta_{iso}(k) &=&\frac{\delta n_{B}^{iso}}{n_{B}}
       \frac{\Omega_{B}}{\Omega_{t}}, \\
        &=&2 \cot 2\theta_{k}
       \frac{H(t_{k})}{\sqrt{2 k^{3}} \phi(t_{k})}
       \frac{\Omega_{B}}{\Omega_{t}},
\end{eqnarray}
where $\Omega_{B(t)}$ is the density parameter of baryons (total matter),
$t_{k}$ is the time when the given mode crosses the Hubble radius during
inflation. On the other hand, both the AD field and the inflaton generate
the adiabatic fluctuation given by
\begin{eqnarray}
       \delta_{ad}(k) &=& \frac{2}{3}
       \left(\frac{k}{a(t)H(t)}\right)^{2}
       \Phi_{H} \\
       &=& \frac{2}{3}
       \left(\frac{k}{a(t)H(t)}\right)^{2}\times
       \frac{2}{3}\frac{1}{M^{2}}\left(\frac{V(\sigma)}{V'(\sigma)}
         \delta \sigma+\frac{V(\phi)}{V'(\phi)}
         \delta \phi \right)\\
       &=& \frac{2\sqrt{2}}{9}
       \frac{\sqrt{k}H(t_{k})}{a(t)^{2} H(t)^{2}M^{2}}
       \left(\frac{8 \pi^{2}}{g^{2}} \sigma(t_{k})
       +\frac{\phi(t_{k})}{2d-2} \right),
\end{eqnarray}
where $t$ is an arbitrary time. To compare these two types of fluctuations,
it may be natural to consider the ratio of the power spectra at horizon
crossing, i.e., $k^{-1} a(t) =H(t)^{-1}$, which is
written as~\cite{ksy}
\begin{equation}
      \label{eq:ratio}
      \alpha_{KSY} = \left. \frac{P_{iso}}{P_{ad}} \right|_{k=a(t)H(t)} =
      \frac{81 g^{4}M^{4}}{256 \pi^{4} \phi(t_{k})^{2} \sigma(t_{k})^{2}}
      \left(\frac{\Omega_{B}}{\Omega_{t}}\right)^{2}
      \cot^{2} 2\theta_{k}.
\end{equation}
It is also useful to consider the ratio $\alpha$ of the present power
spectra~\cite{kanazawa}. The present power spectrum can be described as
\begin{equation}
       P (k)=P_{ad}+P_{iso}=
       A_{ad} k T_{ad}^{2}(k)+A_{iso} k T_{iso}^{2}(k),
\end{equation}
where the transfer functions ($T_{ad,\, iso}$) are normalized
as $T(k \rightarrow 0)=1$. 
The ratio $\alpha$ is defined as
\begin{equation}
       \alpha \equiv \frac{A_{iso}}{A_{ad}}.
\end{equation}
Actually, $\alpha$ is related to $\alpha_{KSY}$~\cite{kanazawa}  by
$\alpha =\left( \frac{4}{27} \right)^{2} \alpha_{KSY}$. From
Eqs.~(\ref{eq:efolding}), (\ref{eq:upper}), and (\ref{eq:ratio}),  we
obtain a lower limit for $\alpha$ as
\begin{eqnarray}
      \alpha &=&
      \frac{g^{4}M^{4}}{144 \pi^{4} \phi(t_{k})^{2} \sigma(t_{k})^{2}}
      \cot ^{2} 2 \theta_{k}
      \left(\frac{\Omega_{B}}{\Omega_{t}} \right)^{2}, \\
      &>&     \alpha_{c},
\end{eqnarray}
where $\alpha_{c}$ is given by
\begin{eqnarray}
       \alpha_{c} &=& \left\{
       \begin{array}{ll}
           \ds{ \frac{\sqrt{30}}{72\pi^{2}}g \lambda \cot^{2}2\theta_{k}
           \left(\frac{\xi}{M} \right)^{-2}N_{k}^{-1}
           \left(\frac{\Omega_{B}}{\Omega_{t}} \right)^{2}} & \ds{d=4}\\
           \ds{ \left(\frac{5}{72} \right)^{\frac{1}{4}}
           \frac{g^{\frac{3}{2}}
           \lambda ^{\frac{1}{2}}}{12 \pi^{2}}\cot^{2}2\theta_{k}
           \left(\frac{\xi}{M} \right)^{-1}N_{k}^{-1}
           \left(\frac{\Omega_{B}}{\Omega_{t}} \right)^{2}} & \ds{d=6}
       \end{array}
       \right..
\end{eqnarray}
If we take $N=50, \, \Omega_{B} =0.03h^{-2}, \, \Omega_{t}=0.28,$
and $ h=0.8$,\footnote{%%
      These are best-fit values of model(10) in~\cite{ekv}.
      The fit was done for the data from Boomerang and MAXIMA-1.}
this lower limit is approximately given  by
\begin{eqnarray}
       \label{eq:lower}
       \alpha_{c} &\simeq &\left\{
       \begin{array}{ll}
           \ds{0.52g\lambda\cot^{2}2\theta_{k}}
           & \ds{d=4} \\
           \ds{8.4\times 10^{-4} g^{\frac{3}{2}} \lambda
           ^{\frac{1}{2}}\cot^{2}2\theta_{k}} & \ds{d=6}
       \end{array}
       \right..
\end{eqnarray}
If $g\simeq \lambda \simeq $0.1 and $\theta \sim O(1)$,
then $\alpha_{c}\sim 1.1 \times 10^{-3}$ for the $d=4$ case.
According to \cite{ekv}, the best-fit value\footnote{%%
    $\alpha_{EKV}$ defined as Eq.(6) in~\cite{ekv} is related to our
   definition of $\alpha$ as follows.
   $$ \alpha = \frac{\alpha_{EKV}}{36(1-\alpha_{EKV})}.$$
    We adopt the model (10) in \cite{ekv}, because
    both adiabatic and isocurvature perturbations are
    almost scale-invariant in our model.}
of $\alpha$ is $\alpha_{best-fit} = 2.4 \times 10^{-3}$.
Finally, we have constraints on $g$, $\lambda$, and $\theta$:
\begin{eqnarray}
      \label{eq:d4}
      g \lambda \cot ^{2} 2 \theta_{k} &<&
      4.7 \times 10^{-3} , \,\,  d=4, \\
      \label{eq:d6}
      g^{\frac{3}{2}} \lambda ^{\frac{1}{2}}\cot^{2}2\theta_{k}&<&
      2.9,\,\,  ~~~~~~d=6 .
\end{eqnarray}
Thus, the large coupling constants can be excluded for $d=4$. On the
other hand,  the constraint for $d=6$ ( and  $d\ge 8$) is not severe at
all.  It is  noticed that the value of the AD field during inflation
can be smaller than that used here for some initial conditions because
the AD field may oscillate before inflation. For this case, the 
constraint becomes more stringent.

\section{Conclusion}

In this paper we have considered the adiabatic and isocurvature
fluctuations of the AD field in the D-term inflation model, and
have found that there exists an upper limit for AD field due
to its dynamical property and the requirement that an inflation
should occur. The primordial spectrum has been
calculated analytically and numerically, and has been found to be
of the familiar Harrison-Zeldovich type.  While the adiabatic
fluctuations of the AD field do not make any significant contribution,
the isocurvature fluctuations of the AD field can generate baryonic
isocurvature perturbations. The upper bound for the AD field in
turn leads to the lower limit for isocurvature fluctuation as
Eq.(\ref{eq:lower}). Taking account of the observational constraints
on isocurvature perturbations from Boomerang and MAXIMA-1,
we had interesting constraints on some combinations of $g$,
$\lambda$, and $\theta$, especially in the case of $d=4$.

\vspace*{0.5cm}
{\sl Acknowledgment:} We thank Masahide Yamaguchi and Tsutomu Yanagida
for useful discussion. This work is supported by the Grant-in-Aid
 for Scientific Research from the Ministry of Education, Science,
 Sports, and Culture of Japan, Priority Area
 ``Supersymmetry and Unified Theory of
Elementary  Particles'' (No.\ 707).

\end{document}